\documentclass[a4paper,11pt]{article}
\usepackage{pos}
\usepackage{amsmath}
\usepackage{bm}

\title{Smeared spectral functions from lattice QCD: opportunities and challenges}
\ShortTitle{Smeared spectral functions from lattice QCD}

\author*[a,b]{Shoji Hashimoto}

\affiliation[a]{Theory Center, Institute of Particle and Nuclear
  Studies, High Energy Accelerator Research Organization (KEK),
  Tsukuba, Ibaraki 305-0801, Japan}

\affiliation[b]{SOKENDAI (The Graduate University for Advanced Studies),
  Tsukuba, Ibaraki 305-0801, Japan}

\emailAdd{shoji.hashimoto@kek.jp}

\abstract{
  A broad class of physical observables can be expressed in terms of
  smeared spectral functions.
  Important examples include the hadronic vacuum polarization
  contribution to the muon $g-2$, hadronic $\tau$ decays, inclusive
  semileptonic $B$ and $D$ decays, and amplitudes for rare processes
  such as $B\to K^{(*)}\ell^+\ell^-$.
  In recent years, a variety of methods have been proposed to
  reconstruct or constrain such spectral functions from Euclidean
  correlators and related inputs. 
  However, the inverse problem is intrinsically ill-posed, making it
  difficult to obtain fully reliable, model-independent results with
  controlled uncertainties.
  In this contribution, I summarize recent developments, discuss the
  opportunities offered by smeared spectral observables, and highlight
  the main challenges that remain.
}

\FullConference{The 43rd International Symposium on Lattice Field Theory (Lattice 2026)\\
July 26 to August 1, 2026\\
University of Maryland, College Park, USA\\}

\begin{document}
\begin{flushright}
  KEK-CP-0413
\end{flushright}

\maketitle

\section{Introduction}

The reconstruction of spectral functions from Euclidean lattice
correlators is a long-standing problem.  The most familiar example is 
the hadronic $R$ ratio for $e^+e^-\to$ hadrons.  Its experimental line
shape contains narrow resonances, thresholds, and broad structures
across the light, charm, and bottom regions.  From lattice QCD one
computes Euclidean correlation functions, whose relation to the
real-energy spectral function is given by a Laplace transform, and the
reconstruction of the local function $R(s)$ from such data is an ill-posed
inverse problem.  

The point of view emphasized here is that many phenomenologically
relevant quantities do not require a pointwise spectral function.
They are instead integrals of the spectral function with a given
weight function (or a {\it kernel}).  Such a smeared spectral function
may be considered either as a compromise forced by the Euclidean
formulation or as a built-in feature of the observable itself.  
For example, the hadronic-vacuum-polarization (HVP) contribution to the muon $g-2$
itself defines a smeared quantity.
Other cases, such as the inclusive semileptonic decay rates or
amplitudes defined with an $i\epsilon$ prescription, can also be
viewed as smeared spectra.
In these cases, the corresponding kernel must often be further smeared
to regulate a potentially singular or sharply varying functional form.
Thus, {\it smearing} plays two distinct roles: it may either define
the physical quantity itself, or serve as a regulator introduced to
make the reconstruction possible. 

In the following, I review the possible applications of the smeared
spectrum ({\it opportunities}) and the available methods and their
limitations ({\it challenges}).
The main message is not that smearing solves the inverse problem once
and for all. 
Rather, smearing changes the question.
Instead of asking for a pointwise spectral function, one asks which
weighted integrals of the spectral function are accessible with the
available lattice data.
The answer depends both on the physical spectrum and on the achievable 
resolution of the reconstruction.

After stating the problem in Sec.~\ref{sec:problem}, we start from 
Sec.~\ref{sec:Opportunities} which surveys the range of physics
applications that can be cast in this form.
Sec.~\ref{sec:Challenges} turns to the reconstruction problem itself
and the sources of its difficulty,
and Sec.~\ref{sec:Concluding_Remarks} collects the resulting lessons for how
small a smearing width or regulator can realistically be achieved. 

We are not able to cover the finite-temperature spectral functions,
although the same inverse problem occurs there.

\section{The problem}
\label{sec:problem}
Let us define the problem. We consider a Euclidean correlator
\begin{align}
  C_{AB}(t) = \langle F|O_A(t)O_B(0)|I\rangle
  = \int_0^\infty\! dE\,\rho_{AB}(E)\,e^{-Et},
  \label{eq:laplace}
\end{align}
where the initial and final states, $|I\rangle$ and $|F\rangle$
respectively, are assumed to be created from the vacuum by some
operator insertions and taking a large Euclidean time separation.
The spectral density may be written as
\begin{align}
  \rho_{AB}(E) = \sum_n 
  \langle F|O_A|n\rangle\langle n|O_B|I\rangle\,
  \delta(E-E_n).
  \label{eq:rho}
\end{align}
The target of the calculation is often a smeared functional
\begin{align}
  S_K[\rho_{AB}] = \int_0^\infty\! dE\,K(E)\rho_{AB}(E),
  \label{eq:smeared}
\end{align}
with a chosen kernel $K(E)$.
If $K(E)$ can be approximated by a linear combination of
exponentials,
\begin{align}
  K(E) \simeq g_0+g_1 e^{-E}+g_2 e^{-2E}+g_3 e^{-3E}+\cdots,
  \label{eq:kernel_expansion}
\end{align}
then the smeared spectrum can be obtained from the Euclidean lattice data as
$S_K[\rho_{AB}]\simeq \sum_t g_t C_{AB}(t)$, since the energy in
(\ref{eq:kernel_expansion}) can be promoted to the Hamiltonian
operator $\hat{H}$ as $e^{-\hat{H}t}$ in the time evolution of the correlator.

It is useful to keep in mind from the outset that smearing plays two
conceptually distinct roles in what follows.
In some cases the kernel is the observable --- the smearing width is
fixed by the physical definition of the quantity, as in the HVP
contribution to the muon $g-2$.
In other cases, smearing is introduced as a regulator to tame an
intrinsically sharp kernel, such as a sigmoid function or an
$i\epsilon$ prescription, and must eventually be removed by an
extrapolation $\sigma\to 0$ or $\epsilon\to 0$.
Both situations reduce to the same reconstruction problem, but the
physical question and the requirement differ between them.
This distinction recurs throughout Sec.~\ref{sec:Opportunities} and is
discussed further. 

Thus, a large class of smeared-spectrum calculations can be formulated
as linear reconstruction problems.
The difficulty is controlled by the smoothness of $K(E)$.
Smooth kernels can be represented by relatively few Euclidean-time
modes, while sharp kernels, such as step functions, delta functions,
or the unregulated $i\epsilon$ prescription, 
require high-resolution information that is hard to reconstruct.

\section{Opportunities --- physics applications}
\label{sec:Opportunities}
In this section, I survey which problems can be formulated as
smeared-spectrum reconstruction problems.
They fall into three classes: inclusive rates, energy-filtered matrix
elements, and amplitudes requiring an energy integral.
I do not attempt to cover applications to parton distribution
functions and related structure functions.
The HVP contribution to the muon $g-2$ is perhaps the best-known example
of the observable with a built-in kernel, but it does not require the
inverse-Laplace reconstruction.

\subsection{Smeared $R$ ratio and inclusive hadronic $\tau$ decays}
A first-principles determination of the smeared $R$ ratio is a useful
benchmark for the methods to reconstruct the smeared spectrum.
The ETMC collaboration initiated such work by defining a Gaussian
smearing kernel with which the experimentally measured $R(s)$ is
convoluted;
the same quantity is obtained from the lattice correlator
using one of the linear reconstruction methods, {\it i.e.} that of
Hansen, Lupo and Tantalo (HLT)~\cite{Hansen:2019idp}.
This provides an alternative way to test QCD: one does not try to
reproduce every resonance peak, but rather to measure a
resolution-limited observable that is well defined on both the
experimental and lattice sides.

The result is encouraging because reasonable agreement is observed.
At the same time, the smearing width used in the available studies is
much broader compared with resonance structures in the light-quark
region. The limitation of the lattice methodology is discussed later.

Inclusive hadronic $\tau$ decay provides another example of a smeared
spectral function. In this case, the smearing kernel is given by the
phase space integral of the decay $\tau\to\nu_\tau X_{ud}$. Schematically,
the decay rate is written as 
\begin{align}
  \Gamma^{(\tau)} \sim
  \int_0^{m_\tau}\frac{d\omega}{\omega^3}
  \left(1-\frac{\omega^2}{m_\tau^2}\right)
  \left(1+2\frac{\omega^2}{m_\tau^2}\right)
  \omega^2\rho_T(\omega), 
\end{align}
where $\rho_T(\omega)$ represents a transverse part of the spectral
function defined for the weak-current HVP.
The smooth part of the kernel is a polynomial function of $\omega$,
which is approximated by a polynomial of $e^{-\omega}$.
The upper limit of the integral can be implemented by a step function;
its sharp step is replaced by a smooth sigmoid, {\it e.g.}
$\theta_\sigma(x)=1/(1+ e^{-x/\sigma})$, 
so that the approximation can be controlled and the limit $\sigma\to0$
can be studied.
Such a calculation has been performed \cite{Evangelista:2023fmt} and the
result can be used to determine $|V_{ud}|$ and $|V_{us}|$.

\subsection{Inclusive semileptonic decays}
Inclusive semileptonic decays of heavy mesons have a similar structure
and are particularly interesting because of their connection with the
inclusive--exclusive tensions in CKM phenomenology.

The total rate can be written in terms of a hadronic spectral
density.\footnote{Closely related ideas have also been proposed for
inclusive neutrino--nucleon scattering~\cite{Fukaya:2020wpp}.}
For example, in the schematic notation used in the talk, the $B$-meson
inclusive semileptonic decay rate is written as
\begin{align}
  \Gamma \propto \int_0^{\bm{q}^2_{\rm max}}d\bm{q}
  \int_{\sqrt{m_D^2+\bm{q}^2}}^{m_B-\sqrt{\bm{q}^2}} d\omega\,
  K(\omega;\bm{q}^2)
  \langle B(\bm{0})|\tilde{J}^\dagger(-\bm{q})
  \delta(\omega-\hat{H})\tilde{J}(\bm{q})|B(\bm{0})\rangle .
  \label{eq:inclusive_b}
\end{align}
The matrix element including $\delta(\omega-\hat{H})$ can be viewed as a
spectral density, which depends on the hadronic energy $\omega$ and
spatial momentum $\bm{q}$.

The kernel function $K(\omega;\bm{q}^2)$ is given by the phase-space
integral for the semileptonic kinematics, and is essentially a
Heaviside function that suddenly vanishes at the kinematic endpoint
$m_B-\sqrt{\bm{q}^2}$. Treating $K(\omega;\bm{q}^2)$ as a smearing
kernel and applying the approximation in terms of $e^{-\omega t}$ was
the basic idea enabling the lattice calculation of inclusive decay
rates \cite{Gambino:2020crt}, and it has been numerically tested
extensively \cite{Gambino:2022dvu,Barone:2023tbl,Kellermann:2025pzt}.

The non-smoothness of the kernel function poses a problem for the
approximation, {\it i.e.} the ill-posed inverse problem returns.
Thus, an artificial smearing is introduced, for example by replacing the
Heaviside function with a smoothed sigmoid, as in the calculation of
the hadronic $\tau$ decay\footnote{In fact, it was introduced first for
the inclusive semileptonic decays \cite{Gambino:2020crt}.}. It should
not be confused with the smearing originating from the phase-space
integral; the artificial smearing must eventually be extrapolated away.

Recent applications to inclusive $D_s$ decays have reached a stage to
quote the decay rate with quantitative error estimates
\cite{DeSantis:2025yfm,DeSantis:2025qbb,Kellermann:2025pzt,Kellermann:2026sgp}. 
The dependence on the recoil kinematics is, however, important.  Near
the kinematic endpoint (large $\bm{q}^2$), the gap between the lower
and upper limits of the $\omega$ integral gets narrow, and the
kernel function becomes more localized. The truncation error of the
approximation or the extrapolation $\sigma\to 0$ is then more
delicate.
Whether a simple extrapolation linear in $\sigma^2$ is justified
depends on the details of the (smeared) 
kernel and the actual physical spectrum near the endpoint. For the
precision achieved so far, such error is not yet the dominant source
of uncertainty \cite{Kellermann:2025pzt}.
As statistical precision improves in the future, the issue should be
inspected again.

For $B$ meson decays, direct simulations are more difficult because
larger recoil momenta lead to noisier correlators and larger
discretization effects.
A promising strategy is to combine nonperturbative lattice
calculations at masses around charm with perturbative inputs and
heavy-quark scaling toward $m_Q\to\infty$ \cite{DeSantis:2026iuu}.
This may be a step toward addressing the inclusive--exclusive puzzle
from first principles, although not yet a complete solution. 

\subsection{Maiani-Testa problem and energy filters}
The Maiani-Testa theorem highlights a basic limitation of Euclidean
correlation functions for extracting infinite-volume scattering
amplitudes~\cite{Maiani:1990ca}.
For example, a Euclidean three-point function with a current and two pion
interpolating fields contains intermediate $\pi(\bm{q})\pi(-\bm{q})$
states of various relative 
momentum $\bm{q}$ but zero total momentum.
At large Euclidean time separations, only the lowest energy states
survive, and in particular the $\bm{q}=0$ state is directly accessible,
{\it i.e.} the time-like pion form factor at the two-pion threshold as
well as the scattering length from a subleading time dependence.

This limitation can be remedied if a Heaviside function
$\Theta(\hat{H}-E_0)$ can be inserted in the matrix element so that
states below a given threshold $E_0$ are discarded, {\it i.e.} a
high-pass filter.
It may also be implemented with a sigmoid function to ease the
approximation.
For the vacuum-to-$\pi\pi$ matrix element, this is the proposal of
Bruno and Hansen~\cite{Bruno:2020kyl}. 
The filtered matrix element can isolate a desired energy region and
relate the result to the time-like form factor and known functions.

Low-pass filters can similarly be used to enhance ground-state
contributions, and energy moments can focus on a state or region of
interest \cite{Bulava:2023brj}.

\subsection{Hadronic amplitudes from spectral functions}
A further class of applications is the construction of hadronic
amplitudes from spectral functions.
In the approaches of Bulava and Hansen~\cite{Bulava:2019kbi},
followed by Frezzotti and collaborators~\cite{Frezzotti:2023nun},
one first defines a Euclidean correlator and an associated spectral
density, then constructs an amplitude as
\begin{align}
  A(E) = \lim_{\epsilon\to 0} \int_0^\infty\!\frac{dE'}{2\pi}\,\frac{\rho(E')}{E'-E-i\epsilon}.
  \label{eq:ieps_kernel}
\end{align}
It probes the spectral function around a specified energy $E$, and can
be considered a smeared spectrum. 
For example, the decay amplitude for $D_s\to\ell\nu\gamma^*$ is
considered for values of $\gamma^*$ energy accessible by
$\gamma^*\to\ell^+\ell^-$.
The expected intermediate states include $K\bar{K}$ and other
multi-particle states. 
A finite $\epsilon$ renders the problem accessible from Euclidean data;
otherwise the imaginary part represents a $\delta$-function
$\delta(E'-E)$, which brings the problem back to the local spectral
function.

The physical amplitude requires the limit $\epsilon\to 0$. 
How this limit is approached depends on the process and its
kinematics.
If the physical amplitude is sensitive to structures narrower than the
achievable $\epsilon$, the extrapolation to $\epsilon=0$ requires
additional physics input on the analytic structure of the
amplitude, including nearby thresholds, poles and coupled-channel
effects. 

Neutral $D^0$-$\bar{D}^0$ mixing is an interesting application.
The short-distance box contribution, which is directly accessible in 
lattice calculations, is expected to be too small to explain the
experimentally observed mixing.
The long-distance intermediate states dominate, but are difficult to
calculate on the lattice.
In a schematic representation, one has to calculate the absorptive
part of the mixing amplitude:
\begin{align}
  \Gamma_{12}\sim \sum_f A(D^0\to f)^* A(\overline D{}^0\to f),
  \label{eq:d_mixing}
\end{align}
where the possible final states include $\pi\pi$, $\pi K$,
$K\bar{K}$, $\eta\eta$, $\pi K^*$, $\rho K$, and many others.
The important restriction is that one has to isolate states whose 
total energy matches the $D$-meson mass.
This makes the problem sensitive to the achievable energy resolution. 

A strategy for lattice calculation has been proposed in which the
relevant long-distance contribution is extracted from 
Euclidean correlation functions through the smeared spectrum
reconstruction \cite{DiCarlo:2025mnm}. 
It includes the integral with the kernel $1/(E'-E-i\epsilon)$;
the physical target requires a limit $\epsilon\to 0$,
and the reliability of that extrapolation depends on both the
reconstruction resolution and the hadronic spectrum. 

These examples are important because they show that smeared spectral
functions are not limited to inclusive rates.
They may also provide access to long-distance contributions to
flavor-changing amplitudes, where sums over many intermediate states
are necessary. 

\subsection{A big challenge: $B\to K^{(*)}\ell^+\ell^-$}
Rare semileptonic $B$ decays offer another compelling but difficult
target.
Deviations in observables such as $B\to K^{(*)}\ell^+\ell^-$ from
their Standard Model predictions are often discussed as potential
signs of new physics,
but the interpretation depends on long-distance hadronic effects,
especially charm-loop and rescattering contributions.
Factorization of the amplitude does not fully describe the
observed spectra, and a fully non-perturbative calculation of the
amplitude is extremely challenging (for discussions, see for example,
\cite{Lyon:2014hpa,Altmannshofer:2026cwk}).

Following the strategy discussed above, one may write a contribution
to the amplitude schematically as (\ref{eq:ieps_kernel}) with $E=m_B$
and 
\begin{align}
  \rho(E) = \langle K(-\bm{q})|\tilde{J}_{\rm em}(\bm{q})
  \delta(\hat H-E) {\cal O}_{1,2}(0)|B(\bm{0})\rangle .
  \label{eq:bkll_rho}
\end{align}
Here, the weak effective operator ${\cal O}_{1,2}$ induces the
transition $b\to s c\bar{c}$, and the electromagnetic current
$J_{\rm em}$ transforms the $c\bar{c}$ pair to $\gamma^*$ and then to
a lepton pair.
The hadronic intermediate state contains the (valence) quark content
$s\bar{d}c\bar{c}$, which is in the hadronic language $\psi K$,
$D_s^*D$, $\cdots$, and their rescattering may induce complicated
hadronic effects. 

A pilot lattice study has been carried out in the small-recoil
region~\cite{Frezzotti:2025hif}.
The calculation uses finite $\epsilon$, with values of order
$1.5$--$3$~GeV, and the limit $\epsilon\to0$ remains to be taken.
To be phenomenologically useful, the lattice calculation must resolve or
constrain the relevant long-distance structures in the charm region.
This may require a substantially smaller $\epsilon$ than is
currently accessible.
This point will be revisited in Sec.~\ref{sec:Concluding_Remarks}.

As anticipated in the introduction, the preceding examples illustrate
the two distinct roles smearing can play, ranging from the
physics-fixed kernel of the HVP-type observables to the regulator-type
smearing required for inclusive decays and hadronic amplitudes.

This leads to two crucial questions.
The first is physical: how small must $\sigma$ or $\epsilon$ be in
order for the limiting procedure to be reliable for the observable of
interest?
The second is technical: how small can $\sigma$ or $\epsilon$ be
implemented with the available Euclidean data?
These questions have to be considered separately.
The first depends on the physical spectrum and the kinematics.
The second is controlled by the inverse problem and by the
signal-to-noise properties of the correlator. 
The next section focuses on the reconstruction methods and their
limitations.

\section{Challenges --- spectral reconstruction}
\label{sec:Challenges}
A large number of methods have been proposed for reconstructing
spectral functions or smeared spectral observables.
My list is likely not comprehensive, but in the field of lattice QCD
(mainly zero-temperature applications\footnote{Admittedly the references
related to the reconstruction of parton distribution functions are
largely missing.}) it includes
maximum entropy methods \cite{Asakawa:2000tr},
Bayesian approaches \cite{Burnier:2013nla,DelDebbio:2021whr},
machine-learning methods \cite{Kades:2019wtd,Chen:2021giw,Wang:2021jou},
Gaussian processes \cite{Horak:2021syv},
Backus--Gilbert methods \cite{Hansen:2017mnd},
sparse modeling \cite{PhysRevE.95.061302,Itou:2020azb},
and Gaussian-quadrature methods~\cite{Aliberti:2026kuq}.
Methods that put more direct emphasis on smeared spectra include
HLT~\cite{Hansen:2019idp},
Chebyshev approximations \cite{Bailas:2020qmv},
conformal-map or Nevanlinna-Pick approaches
\cite{Bergamaschi:2023xzx,Abbott:2025snz,Fields:2025glg},
Bayesian reinterpretations of HLT \cite{DelDebbio:2024lwm},
and recent machine-learning or operator-learning formulations
\cite{Buzzicotti:2023qdv,DeSantis:2026vqg}.
An earlier review by Jay provides useful context \cite{Jay:2025dzl}.

Despite differences in detail, the broad consensus would be:
(1) local spectral reconstruction is indeed ill-posed, and no method
has produced a general, model-independent solution to evade the
ill-posedness, and 
(2) smeared spectra can be reconstructed with controlled errors in
favorable situations, but the details of the kernel, covariance,
{\it etc}. matter.
In short, these methods are useful, but they do not magically remove
the underlying ill-posedness.
As discussed in Sec.~\ref{sec:lenses}, this is because different
reconstruction methods amount to different choices of basis and
criterion to determine corresponding coefficients, but they are 
applied to the same underlying data.
The choice of method cannot by itself increase the information
available from the correlator. 
Let us first discuss why the spectral reconstruction is so hard.

\subsection{Why is it so difficult?}
The singular-value decomposition (SVD) of the Laplace kernel gives a simple
explanation of the
difficulty~\cite{Wang:2021jou,Tsuji:2026zku}.\footnote{For the use of
the SVD basis in the spectral reconstruction, see also
\cite{Rothkopf:2020qqt,Lupo:2026vdj}.} 
On discrete sets of Euclidean times $t_i$ and energies $\omega_j$, the
kernel $e^{-\omega t}$ can be decomposed into orthonormal basis
vectors and singular values:
\begin{align}
  e^{-\omega t} = \sum_l U_l(t) \sigma_l V_l(\omega).
\end{align}
The (real and positive) singular values $\sigma_l$ listed in
descending order (with an integer label $l$) decrease exponentially as
$l$.
The rate depends on the choice of $t_i$ and $\omega_j$, but generally
the modes of large $l$ are strongly suppressed in the Euclidean
correlator, and their reconstruction is unstable or even impossible
once statistical noise is included.
(Even without statistical noise, modes with $\sigma_l$ below 
double-precision arithmetic resolution cannot be reconstructed in practice.)

As one can see by inspecting the functional form of the basis
functions $\{U_l(t)\}$ and $\{V_l(\omega)\}$, the high-mode (large
$l$) components correspond to fine structure in energy. 
In logarithmic variables, a mode of order $l$ resolves structures
roughly on the scale $\Delta\ln\omega\sim 1/l$.
If the corresponding singular value is already below the precision of
the correlator, this fine structure is effectively lost before any
reconstruction method is applied. 

In the SVD basis, $\{U_l(t)\}$ and $\{V_l(\omega)\}$, the smeared
spectrum can be reconstructed as
\begin{align}
  \int_0^\infty\!d\omega\,\rho(\omega) K(\omega) = \sum_l \rho_l K_l,
\end{align}
where
\begin{align}
  \rho_l = \frac{1}{\sigma_l}\sum_{t_i} U_l(t_i)C(t_i) \quad , \quad
  K_l = \sum_{\omega_j} K(\omega_j) V_l(\omega_j).
\end{align}
The correlator components associated with high modes are suppressed by
$\sigma_l$, and its contribution $\rho_l$ is reproduced by multiplying
$1/\sigma_l$, which is the source of the large noise of the
reconstructed spectrum because the noise is also enhanced by this
factor. 
The smearing kernel $K(\omega)$ is also expanded in the SVD basis, and
its contribution to large $l$ modes, $K_l$, is suppressed when $K(\omega)$ is
smooth and does not contain the high-frequency components.
Since the smeared spectrum is a sum of the product $\rho_l K_l$, a
stable result is obtained only when $K_l$ decreases faster than the
growth of the noise in $\rho_l$.\footnote{The growth rate of the noise
  in $\rho_l$ itself is a non-trivial question, but as a first
  approximation one could consider the case where the noise in $C(t)$ is
  constant as a function of Euclidean time. When the noise grows
  according to the Parisi-Lepage mechanism, one can consider a product
  $e^{m_\pi t}C(t)$ instead, so that the noise stays constant.}
The wider the smearing width, the better the achievable precision for
the smeared spectrum, as we anticipated.


\subsection{Mellin transform and logarithmic resolution}
In the limit of continuum $t$ and $\omega$ in the range
$[0,\infty]$, the singular value decomposition of the Laplace kernel
approaches the Mellin transform, which is extensively discussed in
\cite{Bruno:2024fqc,Giusti:2026mcy}. 
To be explicit, it is written as
\begin{align}
  e^{-\omega t} = \int_0^\infty\! ds\, |\lambda_s|
  \left(u_s^+(\omega)u_s^+(t)-u_s^-(\omega)u_s^-(t)\right),
\end{align}
where the ``singular values'' are given as
$|\lambda_s|=\sqrt{\pi/\cosh(\pi s)}$ and the basis functions are
$u_s^\pm(x) =
\renewcommand{\arraystretch}{0.6}
\begin{array}{c}\cos\\\sin\end{array}\!\!\!
\left( s\ln(x) - \theta(s)/2 \right)/\sqrt{\pi x}$.
These modes are Fourier-like functions in logarithmic variables, and
their singular values are exponentially suppressed at high frequency.
Since no lattice spacing is involved in this continuum representation,
the Mellin basis provides a natural language for discussing smearing
functions and the information content of Euclidean correlators. 

The practical consequence is that the energy resolution improves only
logarithmically with correlator precision.
If $\Delta$ denotes the scale of the relative error in the Euclidean
correlator data, a rough estimate of the possible resolution for the
energy spectrum gives
\begin{equation}
  \frac{\Delta\omega}{\omega} \sim \frac{\pi^2}{2\ln(1/\Delta)}.
  \label{eq:log_resolution}
\end{equation}
With currently available correlator precision, say
$\sim 10^{-5}$ or $10^{-8}$,
a realistic resolution may be at the level of $20$--$40\%$ in
energy.
This estimate should not be read as a rigorous theorem, but a
guide to what can be expected without additional physics input. 
For better energy resolution, which may be needed for some physics
applications, exponentially better correlator precision is required,
which is not achievable by brute-force increase of statistics.

\subsection{Many lenses, same aperture}
\label{sec:lenses}
In general, the reconstruction problem can be viewed as a choice of
basis times a criterion for choosing coefficients.
Bases include simple powers of $z=e^{-\omega}$, Chebyshev polynomials
of $z$, SVD or Mellin bases, and many variants.
Criteria include direct inversion in a truncated space, approximation
theory (for the Chebyshev approximation), loss minimization with
statistical covariance, maximum entropy, Gaussian processes, machine 
learning, and other regularization prescriptions.\footnote{See, for
example, \cite{Barone:2023tbl} for a test of multiple bases and criteria.}

Changing the basis or the criterion can improve stability, encode
useful prior information, or make the error analysis more transparent.
But, changing the {\it lens} does not create new information.
The accessible subspace is determined by the Euclidean data with their
covariance matrix, the available time range, and the singular values of the
kernel.
This is the sense in which many methods represent different
{\it lenses} looking through the same {\it aperture}.
Additional inputs, as in Bayesian methods and related approaches, can
be viewed as {\it filters} that enhance selected aspects of the image.

The output of various methods can be different if one looks closely,
but the fundamental limitations described above are shared.\footnote{Nevanlinna--Pick
interpolation is nonlinear and has its own analytic
structure~\cite{Bergamaschi:2023xzx,Fields:2025glg}; nevertheless, the
information available from the Euclidean correlator remains limited.} 
The focus should therefore be on the method used to estimate
truncation errors and systematic uncertainties, not only on the
reconstructed central values. 

\subsection{Approximation versus bounds}
Approximation methods discussed so far construct a representation of
the target kernel in terms of the polynomials of $e^{-\omega}$ and
then estimate the truncation error.
The central question is: what is the error caused by the missing high
modes or by the incomplete kernel approximation?
Various strategies have been developed for HLT, Chebyshev expansions,
and related methods.
Among them, the Chebyshev approximation provides a mathematically
solid ground, because the matrix elements
$\langle T_j(e^{-\hat{H}})\rangle$ are bounded by $\pm 1$ (for
appropriately normalized states).
The size of the truncated terms $c_j\langle T_j(e^{-\hat{H}})\rangle$
can thus be rigorously bounded.

A complementary strategy is to compute rigorous bounds from the
outset.
If the spectral function is positive, one may maximize or minimize the
target smeared quantity \cite{Lawrence:2024hjm}. 
In schematic form, one searches over positive $\rho(\omega)$ subject to
$C(t_i)=\int d\omega\,\rho(\omega)e^{-\omega t_i}$,
and maximizes or minimizes $\int d\omega\,K(\omega)\rho(\omega)$.
This is a convex optimization problem, analogous in spirit to bootstrap bounds.
The method can be extended to include statistical noise
\cite{Abbott:2025snz,Abbott:2026wdw,Mutzel:2026vyw}.
Such bounds may be potentially tighter or more conservative than an
approximation-based error estimate, because they give bounds on the
smeared spectrum itself rather than the kernel function.\footnote{One
caveat is that the method relies on the positivity of the spectral 
function, which is not the case for many of the applications we
discussed in the previous section.} 
Explicit benchmark comparisons between approximation and bound
strategies are highly desirable. 

\subsection{Finite-volume effects}
Smeared spectra often represent two-particle or multi-particle states,
so finite-volume effects can be significant.
Individual finite-volume energy levels receive power-like
finite-volume effects, and the finite-volume spectrum is discrete.
As a result, the finite-volume correlator may behave as $1/L^3$ (or
some other powers).
An explicit example for the vector current correlator is found in
\cite{Itatani:2024fpr}, where the correlator scales as $1/L^3$ at 
sufficiently large Euclidean time separation when a few low-lying
discrete energy levels dominate, while the normal exponential
finite-volume effect $\sim e^{-m_\pi L}$ is found for small time
separations. 

A similar scaling can occur for smeared spectra.
A smearing kernel that is too narrow compared with the level spacing
will resolve the finite-volume discreteness, leading to power-like 
finite-volume effects.
On the other hand, for sufficiently broad smearing, finite-volume
effects in smeared spectral densities can be exponentially suppressed.
Recent analyses~\cite{Bresciani:2026kjv} generalize ideas 
from the finite-volume treatment of the hadronic-vacuum-polarization
contribution~\cite{Hansen:2020whp}.
The suppression can be expressed in terms of an effective mass scale, for example
$\sim \exp(-m_{\rm eff}L)$, with 
$m_{\rm eff}=\frac12\,\mathrm{Im}\sqrt{(\omega^*+i\sigma)^2-4m_\pi^2}$,
where $\omega^*$ is the target energy and $\sigma$ is the smearing width.
The smearing width should be large enough to cover more than one
finite-volume level and suppress finite-volume artifacts, but small
enough not to erase the physical structure of interest. 

\section{Concluding remarks}
\label{sec:Concluding_Remarks}
Through the analysis of the reconstruction strategies, one finds an
inherent limitation in the resolution one can achieve for the energy
spectrum and thus a corresponding limitation on the smearing kernel.
With currently available correlator precision, a realistic resolution
can be roughly estimated as $\epsilon\gtrsim(0.2$--$0.4)E$.

This limitation on resolution influences the range of applications and
their achievable precisions. 
For a relatively smooth kernel required for the inclusive hadronic
$\tau$ decay, this does not induce a significant problem.
On the other hand, for the inclusive semileptonic $B$ and $D$ decays a
careful assessment is desired, since how much the kernel resolution
affects the decay rate depends on the kinematic region.

For a kernel of the form \eqref{eq:ieps_kernel} to calculate the
physical amplitudes, the real and imaginary parts probe structures
over a width set by $\epsilon$. 
If the physical amplitude is controlled by structures narrower than
this width, the extrapolation to $\epsilon=0$ becomes non-trivial.
Taking the application to the $B\to K^{(*)}\ell^+\ell^-$ amplitudes as
an example, the $R$ ratio in the charm region illustrates the
issue. 
Experimentally, the spectrum contains a complicated sequence of
charmonium resonances and thresholds.
Coupled-channel analyses of charmonium data find rich pole structures
with widths and separations of order tens to hundreds of
MeV~\cite{Nakamura:2023obk}.
The relevant spectral density may therefore vary significantly on
scales below 100~MeV. 
For hadronic amplitudes such as rare $B$ decays, this suggests that
one may need $\epsilon<$ 100~MeV, or at least a process-dependent
understanding of why a larger $\epsilon$ is sufficient.
For the $D^0-\bar{D}^0$ mixing, detailed spectral information is not
available, and obtaining it is an important subject for future study. 

This conclusion is not meant to discourage the program.
Rather, it emphasizes that the zero-smearing limit is a physics
problem as well as a reconstruction problem.
One must understand the underlying spectrum, intermediate states, and
kinematic regime before deciding how small the regulator must be. 

To conclude, smeared spectral functions open a broad area of new
applications of lattice QCD.  They provide access to inclusive
observables and to hadronic amplitudes that are otherwise difficult to
access.
The examples discussed above include the smeared $R$ ratio, hadronic
$\tau$ decays, inclusive semileptonic decays, energy-filtered matrix
elements, long-distance contributions to $D^0$--$\bar{D}^0$ mixing,
and rare $B\to K^{(*)}\ell^+\ell^-$ amplitudes. 

The challenges are clear.
There is no universal solution to the ill-posed inverse problem other
than accepting some form of smearing or adding controlled physics
input.
There is also no universal solution to the signal-to-noise problem;
the achievable resolution is limited by the precision and time range
of Euclidean correlators.
Hadronic amplitudes may require rather small values of $\epsilon$, and
one must take the underlying spectrum and the relevant
multi-body states seriously. 

The practical question should therefore be phrased as follows.
Instead of asking whether spectral functions can be fully
reconstructed, one should ask whether phenomenologically useful
smeared observables can be constructed with controlled errors.
This is a less ambitious but more productive target.
In favorable cases it is already achievable;
in more challenging cases, especially amplitudes with complicated
intermediate states, the next step may be to combine reconstruction
methods with physics input, whose impact should also be
clarified. 
Recent proposals for rare charm decays provide one example 
of a phenomenological target where this approach may be useful
\cite{Juttner:2026fui}. 

\section*{Acknowledgments}
I thank the organizers of Lattice 2026 and the many colleagues whose
recent work shaped the discussion summarized here.
In particular, I thank Mattia Bruno, John Bulava, Francesca 
Bresciani, Norman Christ, Sarah Fields, Sakura Itatani, Andreas
J\"uttner, Will Jay, Ryan Kellermann, Davide Laudicina, Alessandro
Lupo, Gabriele Morandi, Alexander Rothkopf, Ryutaro Tsuji
for correspondence, discussions, and comments.
This work is supported in part by JSPS KAKENHI Grant Number 22H00138.

\bibliographystyle{JHEP}
\bibliography{refs}

\end{document}